%% file: conference_101719.tex
\documentclass[conference]{IEEEtran}
\IEEEoverridecommandlockouts
\usepackage{cite}
\usepackage{censor}
\usepackage{amsmath,amssymb,amsfonts}
\usepackage{algorithmic}
\usepackage{graphicx}
\usepackage{textcomp}
\usepackage{xcolor}
\usepackage{tablefootnote}
\usepackage{multirow}
\usepackage{flexisym}

\newcommand{\tabref}[1]{Table~\ref{#1}}
\newcommand{\figref}[1]{Figure~\ref{#1}}
\newcommand{\secref}[1]{Section~\ref{#1}}

\def\BibTeX{{\rm B\kern-.05em{\sc i\kern-.025em b}\kern-.08em
    T\kern-.1667em\lower.7ex\hbox{E}\kern-.125emX}}
\begin{document}

\title{Impacts of Students' Academic-Performance Trajectories on Final Academic Success\\
}

\author{\IEEEauthorblockN{Shahab Boumi}
\IEEEauthorblockA{\textit{Industrial Engineering and Management Systems} \\
\textit{University of Central Florida}\\
Orlando, United States \\
{sh.boumi@knights.ucf.edu}}
\and
\IEEEauthorblockN{Adan Vela}
\IEEEauthorblockA{\textit{Industrial Engineering and Management Systems} \\
\textit{University of Central Florida}\\
Orlando, United States \\
adan.vela@ucf.edu}

}

\maketitle

\maketitle
\input{abstract.tex}


\input{introduction.tex}

\input{background.tex}
\input{problemstatement.tex}
\input{methodology.tex}
\input{analysis.tex}
\input{conclusion.tex}

\bibliographystyle{abbrv}
\bibliography{ref}

\end{document}

%% file: abstract.tex
\begin{abstract}

Many studies in the field of education analytics have identified student grade point averages (GPA) as an important indicator and predictor of students' final academic outcomes (graduate or halt). And while semester-to-semester fluctuations in GPA are considered normal, significant changes in academic performance may warrant more thorough investigation and consideration, particularly with regards to final academic outcomes. However, such an approach is challenging due to the difficulties of representing complex academic trajectories over an academic career. In this study, we apply a Hidden Markov Model (HMM) to provide a standard and intuitive classification over students' academic-performance levels, which leads to a compact representation of academic-performance trajectories. Next, we explore the relationship between different academic-performance trajectories and their correspondence to final academic success. Based on student transcript data from University of Central Florida, our proposed HMM is trained using sequences of students' course grades for each semester.  Through the HMM, our analysis follows the expected finding that higher academic performance levels correlate with lower halt rates. However, in this paper, we identify that there exist many scenarios in which both improving or worsening academic-performance trajectories actually correlate to higher graduation rates.  This counter-intuitive finding is made possible through the proposed and developed HMM model.

\end{abstract}

%% file: introduction.tex
\section{Introduction}
Based on the National Student Clearinghouse Research Center, the six-year graduation rate for students who initiated their college education in 2012 is only 58\%, with 42\% of students either halting enrollment or taking longer than six years to graduate \cite{shapiro2018completing}. Halting college is broadly understood to impose irreversible mental, financial, and time losses to students \cite{garibaldi2012college}. %Apart from that, based on results reported by Vitasari et al.  \cite{vitasari2010relationship}, there is a significant correlation between study anxiety and academic performance. In other words, low academic performance has results in high study anxiety and vice versa. As such, many top-notch universities establish financial and educational resources to support students at risk of leaving schools without earning a degree.

To date, numerous reasons have been identified to explain why students choose to halt their education; they include: financial problems, lack of interest in studies, lagged behind in study progression, and inadequate information and guidance \cite{hovdhaugen2009learning}. In many cases, the reasoning is related to academic performance; however, the determination to halt one's education is not a straight line between perceived poor grades and \textit{halting}.  As D. Shippee notes, poor grades may lead to the feeling of depression \cite{shippee2011gpa}, or lead students to shift career goals \cite{hull2005career}.  Instead, low academic performance can be viewed as an underlying factor, predictor, or leading indicator of halting \cite{aghajari2020decomposition}.  Along these lines, many studies have recognized GPA as a fair predictor or leading indicator of students persisting or halting their academic career \cite{asee_peer_34921,ebrahiminejad2019pathways,pappas2016investigating,marbouti2016models,mirzaei2019modeling}. Zahedi et al. \cite{asee_peer_34921} investigated the relationship between graduation rate and numerous variables, including gender, race, transfer status, major, number of terms registered, and cumulative GPA. Their findings suggest that cumulative GPA is one the most powerful measures for predicting graduation. In a similar study, Pappas et al. \cite{pappas2016investigating} showed that students' cumulative GPA is a significant factor in identifying students at the risk of dropout among computer science students. Although cumulative GPA is demonstrated to be a valuable indicator and predictor of a students' academic outcomes, the measure is a summary statistic.  As such, cumulative GPA does not convey information of how a student's GPA evolved over time.  Using cumulative GPA as a predictive measure essentially ignores underlying patterns that could convey meaningful status changes.   

To better understand, model, and characterize students at-risk of halting, we believe it is necessary to mathematically model academic-performance trajectories.  The desire to model academic performance trajectories is driven by the following research question:
\begin{itemize}
\item Does a decrease in academic performance proportionally increase the risk of halting?  \item If a student improves their academic performance does their graduation rate match those students that always performed well? 
\end{itemize}
%Students can be observed to exhibit a variety of academic performance trajectories over their academic careers.  As such, student may end their academic careers by graduating with a range of cumulative GPAs.  Alternatively, students may halt their education, either on their own accord\footnote{We are not implying that a student's decision to halt is willingly, or that the decision to halt is based on academic performance.} or as a result of enforced university policies  (e.g. academic probation leading to disqualification, academic misconduct, etc).  To better understand, model, and characterize students at-risk of halting, we believe it is necessary to mathematically model academic-performance trajectories.  The desire to model academic performance trajectories is driven by the following research question: \textit{Does a decrease in academic performance proportionally increase the risk of halting?}  And furthermore: \textit{If a student improves their academic performance does their graduation rate match those students that always performed well?}. 

This study is aimed at analyzing students' academic-performance trajectories to see if these trajectories are meaningful in answering the research questions above.  To address the research questions, we start by designing a Hidden Markov Model (HMM) to convert complex student academic records to a compact discrete representation we denote as an \textit{academic-performance level}.  The HMM is tuned using anonymized student records of first-time-in-college (FTIC) students.  After applying the HMM, the resulting compact representation of the academic-performance trajectories is grouped into meaningful clusters for analysis.  Groups are compared and contrasted to understand how academic-performance trajectories correlate to halting.  Accordingly, the primary contributions of this paper are: \textbf{(1)} a technique for generating a compact representation of a student's academic-performance trajectory; and \textbf{(2)} an improved understanding of the relationship between academic-performance trajectories and final academic outcomes.

The remainder of the paper is organized as follows: in \secref{sec:problemstatement} we provide a more detailed problem statement, with \secref{sec:methodology} detailing our proposed approach. Next, we provide our findings in \secref{sec:results}. Finally, the paper is concluded in \secref{sec:conclusion}.

%% file: problemstatement.tex
\section{Problem Statement}\label{sec:problemstatement}
\begin{figure}
\centering
\caption{Representation of overall problem and solution methodology.}
\centerline{\includegraphics[width=250pt]{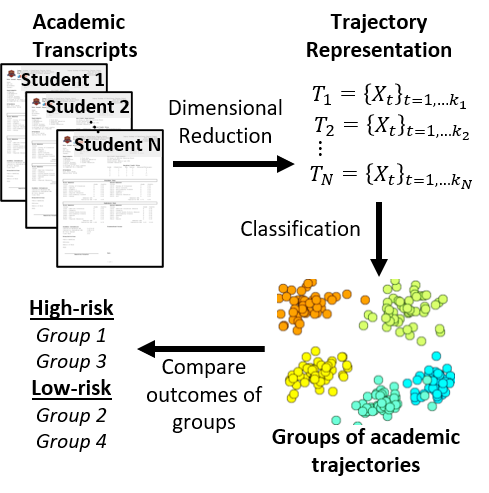}}
\label{fig:problemMethod}
\end{figure}

We consider the problem of mathematically modeling the trajectory dynamics of a student's academic performance.  The goal of addressing this problem is to identify a compact trajectory representation of academic performance that enables the classification and clustering of undergraduate students into meaningful groups.  These groups will then be compared and contrasted to explore the relationship between academic performance trajectories and final educational outcomes, specifically halting.  A pictorial representation of the overall problem and methodology is provided in \figref{fig:problemMethod}.

Here, we start with the notional definition of an academic-performance trajectory to be a temporal sequence of states indicating a student's academic performance.  A frequently encountered category of academic performance trajectories are student transcripts; each semester corresponds to a time-step $t$ in a trajectory $\hat{T}_i$, while courses are taken, and grades received each semester correspond to trajectory states, $\hat{X}_t$.  The challenge with working such data is the dimensionality of the representation - specifically all the combinations of courses and grades in $\hat{X}_t$.  

The goal of this paper is to identify a mapping $\hat{T}_i\rightarrow T_i$ that translates an academic performance trajectory into a lower-dimensional representation $T_i$, where the state $X_t$ at each semester corresponds to an academic-performance level. As an example, consider the case in which each student transcript is first converted to a sequence of grade counts given by the number of A's, B's, C's, D's/F's (combined), and withdrawals they received in a semester.  One such tuple, (1,0,1,1,2), would indicate that in the corresponding semester, a student received the following marks: one A, no B's, one C, one D or F, and two withdrawals.  Over an academic career, a semester-to-semester sequence might look like: 
\begin{equation*}
    (1,0,1,1,2), (2,1,1,1,0),(0,3,1,0,0),\hdots,(3,1,0,0,0).
\end{equation*}  
A mapping of the full trajectory to a compact trajectory representation might be $L_3,L_2,L_2,\hdots,L_1$, where each $L_i$ in the time-ordered sequence corresponds to a specified academic-performance level (e.g. $L_1$ is low, $L_2$ is middle, and $L_3$ is high).  

Unlike prior studies, we explicitly assume the grades a student earns exist within a random process whereby students' course grades are not deterministic.  As an analogy, in an imaginary multiverse, a student could take the same course one-hundred times and earn different scores each time.  Seemingly random processes or disturbances that might affect a student's scores, and ultimately final grade, can be due to external factors (e.g., illness the week of an exam) or internal factors (e.g., grading policies) related to the course.  Accordingly, instead of each academic-performance level, $L_i$ being a straight-forward mapping from the semester GPA, random processes and disturbances that create a noisy GPA signal must be filtered to estimate the academic-performance level.

%This challenge is typically addressed by taking snapshots of academic progress or performance in specific courses (e.g. cumulative GPA after freshman year, course grade in Physics 1B).  As noted however, these types of snapshots do not codify a trajectory, and inherently lose information.  As an alternative, in this paper we propose the use a Hidden Makov Model to encode academic performance trajectories into a compact form, thereby performing a dimensional reduction.

%After applying the dimensional reduction  trajectories are grouped for analysis.  As demonstrated later, the groups are a natural extension of the applyication of the Hidden Markov Model   

%% file: methodology.tex
\section{Methodology}\label{sec:methodology}
In order to map full academic trajectories to a more compact representation, we develop and apply a Hidden Markov Model (HMM).  Application of HMMs to model students' academic behaviors and performance has been documented in several studies  \cite{tadayon2020predicting,boumi2021quantifying2,homsi2008hidden,boyer2011investigating,falakmasir2016data,nguyen2013new}. For example, in work by Boumi and Vela  \cite{boumi2019application,boumi2021quantifying}  the authors use an HMM to investigate the relationship between students' academic performance and enrollment strategies (i.e., the sequence of full-time and part-time enrollment statuses). Through the application of an HMM, students are classified and clustering into three categories based on their enrollment strategy -- full-time, part-time, and mixed enrollment strategy.  Once clustered, the categorizations are demonstrated to correlate with various measures of academic performance (e.g., graduation rate, halting, GPA). In another study by Kaser et al. \cite{kaser2017modeling}, the authors apply an HMM to exam the impact of students' exploration strategies on learning. Their findings suggest that exploration strategy is important to learning outcomes.  Using a similar framework to these studies, we develop and apply an HMM to model and understand the impact of academic-performance trajectories and final educational outcomes. 

In this section, we begin by describing the student data used for this study.  Next, we provide an overview of the Hidden Markov Model used in analyzing students' \textit{academic-performance trajectories}.  Followed is a discussion of the problem formulation and prerequisite steps to encoded academic records and develop the HMM structure. Lastly, we provide the relevant parameters of the trained HMM model.

\input{data}
\input{hmm}
\input{training}

%% file: data.tex
\subsection{Student Data Records} \label{sec:data}
All the findings in this study are based on undergraduate academic records collected by the Institutional Knowledge unit at a large public university University of Central (UCF).  The database of academic records covers the years from 2008 to 2017 and contains course grades for over 100,000 more than 170,000 students. To provide some contextual information regarding UCF, some demographic data is provided in \tabref{tab:demographic_dist}.  As indicated by the  statistics contained in the tables, UCF is a coeducational university with a large fraction of students self-identifying as Hispanic\footnote{As of 2018 UCF was designated to be a Hispanic Serving Institution by the Department of Education.}.  Also unique to UCF is that a large portion of the student body are transfer students, with most coming from community colleges.  

%\begin{table}[htbp]
%\centering
%\caption{Students gender %distribution at UCF over 10 %years}
%label{tab:Gender}
%\begin{tabular}{|c|c|c|} %\hline
%{}&Females&Males\\ \hline
%Percentage&56\% & 44\%\\ %\hline
%\end{tabular}
%\end{table}

%\begin{table}[htbp]
%\centering
%\captioncoeducational{Enrollment type distribution for %different semesters at UCF over 10 years}
%\label{tab:Full-time part time semesters}
%\begin{tabular}{|c|c|c|} \hline
%Semester&Full-time&Part-time\\ \hline
%Fall&73\% & 27\%\\ \hline
%Spring&71\% & 29\%\\ \hline
%Summer&10\% & 90\%\\ \hline
%\end{tabular}
%\end{table}

%\begin{table}[htbp]
%\centering
%\caption{Students admission type distribution at UCF over 10 years}
%\label{tab:admission type}
%\begin{tabular}{|c|c|c|} %\hline
%{}&First-Time-in-College&Tran%sfer\\ \hline
%Percentage&41\% & 59\%\\ %\hline
%\end{tabular}
%\end{table}

%\begin{table}[htbp]
%\centering
%\caption{Students ethnicity distribution at UCF over 10 years}
%\label{tab:Ethnicity}
%\begin{tabular}{|c|c|c|c|c|} \hline
%{}&White&Hispanic&African-Am.&Other\tablefootnote{The other category includes American-Indian, Asian, Native Hawaiian, and Multi-racial ethnicity}\\ \hline
%Percentage&55\% & 24\%&11\%&10\%\\ \hline
%\end{tabular}
%\end{table}

\begin{table}
\centering
\caption{Students demographic distribution at UCF over 10 years}
\label{tab:demographic_dist}
\begin{tabular}{|c|c|} \hline
Students demographic&Percentage\\ \hline
Female&56\%\\
Male&44\%\\  \hline
First-Time-in-College&41\%\\ 
Transfer & 59\%\\ \hline
White&55\%\\
Hispanic&24\%\\
African-Am.&11\%\\
Other\tablefootnote{The other category includes American-Indian, Asian, Native Hawaiian, and Multi-racial ethnicity}&10\%\\  \hline 
\end{tabular}
\end{table}

While transfer students make up the majority of the student body, this study focuses strictly on first-time-in-college (FTIC) students.  This decision is driven by the following reasons: \textbf{(1)} the lack of academic records for transfer students from their previous university; and \textbf{(2)} potentially insufficient time spent at UCF to define a trajectory.  After filtering for FTIC students, the data-set is further restricted to those students that graduated or halted their education\footnote{A student is defined to halt if they do not enroll for three consecutive academic semesters.}.  The remaining data-set contains over $\sim 20,000$ 37,423 unique students and a total of over $\sim 200,000$ 334,177 student-semesters. 

%% file: hmm.tex
\subsection{Representing Academic Performance using a Hidden Markov Model}
\begin{figure}
\centering
\caption{Representation of a simple Hidden Markov Model}
\centerline{\includegraphics[width=250pt]{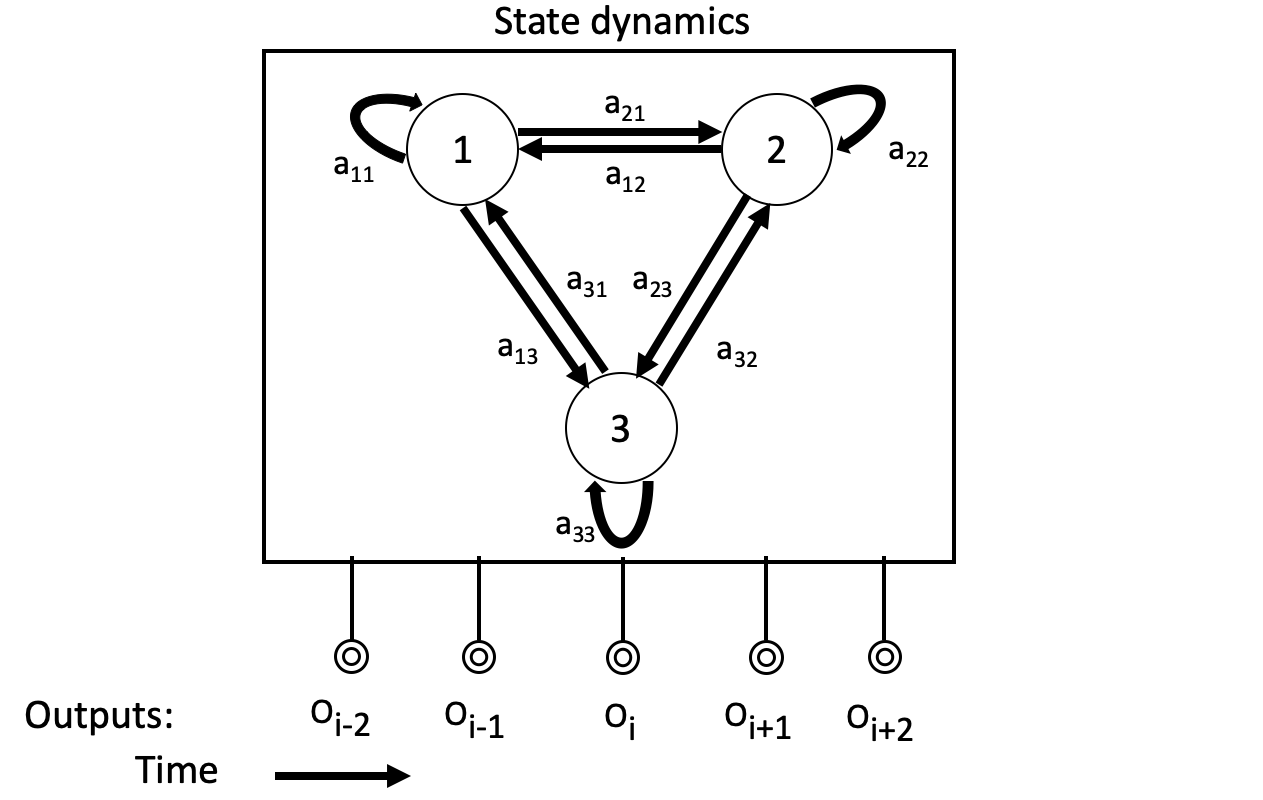}}
\label{fig:HMM}
\end{figure}

As depicted in \figref{fig:HMM}, an HMM represents the dynamics of a system as it moves between operating states or modes (e.g., Modes 1, 2, and 3 in the figure).   When operating within a state or mode, the system generates a mode-related output $O_i$  at each time-step.  For the problem under consideration here, the modes correspond to the academic-performance level of a student, and the observations refer to the student grade counts in any given semester.  Because of the randomness assumption stated in \secref{sec:problemstatement}, the grade counts observed might not map directly to the expected academic-performance level of a student, and as such, we say the mode or academic-performance level of a student is not directly observable from their grades.  Instead, through a series of grade observations, we can estimate or infer the most likely academic-performance level of the student in any semester given the likelihood of such a sequence of mode transitions (i.e., trajectory).

Each Hidden Markov Model is initialized by a set of modes $Q=\{q_1,q_2,\hdots,q_N\}$. Unlike any other types of Markov models in which the modes are observable, in HMMs, modes are hidden; that is, it is unknown at which mode the system dwells at any particular moment. After developing an HMM, one application of an HMM is then to estimate a variable-length sequence of the hidden modes estimates given a series of observations, $O=o_1,o_2,\hdots,o_T$, that are generated from the modes. Within this framework, each HMM is defined by the following parameters: (1) $A=[a_{i,j}]\in\mathcal{R}^{NxN}$, a transition matrix containing the single-step probabilities by which a system moves between modes; (2) $B=b_i(o_t)$, an emission matrix describing the probability of generating each observation $o_t$ when the system is in mode $q_i$; and (3) $\pi_0$, the initial probability distribution for the system at its start.

The first step to build an HMM, known as the \textsc{learning step}, is to estimate the optimal values for the parameters $\lambda=(A, B,\pi_0)$. We use the Baum-Welch algorithm, an expectation-maximization algorithm, to estimate the optimal parameter-set $\lambda^*$. This algorithm takes initial guesses for each of the parameters and iteratively improves the estimations of the parameters by computing the likelihood of any sequence of observation given $\lambda$. After obtaining the optimal $\lambda^*$, the next step is to predict the sequence of hidden modes for a given observation sequence. This step, known as \textsc{decoding step} uses the Viterbi algorithm, which takes observation sequences and $\lambda^*$ as input and returns estimated hidden modes as the output.    

In our case, the first challenge with utilizing an HMM is the dimensionality of the observable outputs, i.e., grade counts.  As described in \secref{sec:problemstatement}, each semester, we represent a student's academic performance according to a tuple with five elements, in which the first, second, and third elements are integers indicating the number of courses with A, B, and C grades. The fourth and fifth elements are variables indicating the number of D or F grades, and how many courses the student has withdrawn. Under the assertion that the grade counts earned each semester are dependent (e.g., the number of A's earned is correlated to the number of B's, C's, etc.), it is reasonable to consider each unique grade-tuple to be a possible observation.  Within the data-set under consideration, over 687 unique grade combinations in a semester have been observed.  The most common grade combination was (4,0,0,0,0), which occurred 5.8\% of the time.  Meanwhile, there are over 20 grade combinations that occur only once, e.g. (1,1,1,0,4).

Given the large number of grade combinations, the wide distribution spread, and the relative sparseness in which many cases occur, it is necessary to perform a partial reduction in the dimensionality of the output space.  Instead of each output being a tuple of unconstrained counts, elements of the grade-tuple are capped. The grade-tuple is adjusted to contain the number of A's, B's, and C's up to a maximum value of 3.  Furthermore, instead of storing the number of D/F and W grades, these values are converted to binary variables.  In this new representation the original grade-tuple $(5,0,2,0,3)$  for a student that had registered for 10 classes would be converted to $(3,0,2,0,1)$.  Now, because students do not typically register for more than 5 classes, and more so, because there is a strong correlation between grades, the impact of clipping is believed to be limited.  After applying the new representation, the total number of unique observations is reduced to 216 grade-tuples.

Within the HMM framework, the next step is to specify a predetermined number of modes for the model; again, each mode corresponds to an academic-performance level.  The inclusion of each additional state increases the required number of parameters used to define the model (i.e., size of $\lambda^*$); it is critical to balance over-fitting (too many parameters) and model coarseness (too few parameters). For a model with $N$ and $M$ outputs, the total number of parameters within an HMM to be tuned is $N(N+M-2)+N$; the tuned parameters correspond to a probability transition matrix, the emission probability for each grade tuple when operating in a state, and the initial state distribution.  Through a combination of managing the number of modes and the number of grade-tuples represented, we can ensure there is sufficient data to tune the parameters.  

Initially, we classify students based on their last semester's cumulative GPAs (CGPA) into eight academic-performance levels. The first level corresponds to students with a CGPA less than 2.0; level 2 includes students with a CGPA between 2.0 and 2.25, level 3 for students with CGPA between 2.25 and 2.5, and so on for other levels. Next, for each academic-performance level, the distribution over grades including A, B, C, DF, and W was computed. Since the grade distributions observed over adjacent academic-performance levels were quite close, a round of aggregation over academic-performance levels was initiated using their pairwise distance. Based on \cite{bhattacharyya1946measure}, the distance between two Multinomial distribution, $\Delta$, is computed by the following equation: 
\begin{equation*}
4\sin^{2}(\Delta/2)=(\sqrt{p_A}-\sqrt{p\textprime_A})+...+(\sqrt{p_W}-\sqrt{p\textprime_W})
\end{equation*}
In the above equation, $p_A$ through $p_W$ corresponds to the probability of grade A through grade W in a given academic-performance level. Given the pairwise distances between academic-performance levels and using the hierarchical clustering method, the eight different academic-performance levels were further grouped into four main levels. We consider the resulting four academic-performance levels as the hidden states for the proposed HMM. The cumulative GPAs associated with these states are shown in \tabref{tab:ClusterCGPA}.

\begin{table}
\centering
\caption{Cumulative GPA for the four academic-performance levels obtained by the hierarchical clustering method}
\begin{tabular}{|c|c|} \hline
Level number&Cumulative GPA\\ \hline
1 & CGPA$\leq$2.5\\ \hline
2 & 2.5$<$CGPA$\leq$3\\ \hline
3 & 3.0$<$CGPA$\leq$3.5 \\ \hline
4 & 3.5$<$CGPA$\leq$4.0\\
\hline\end{tabular}
\label{tab:ClusterCGPA}
\end{table}

%The created hidden Markov model takes a sequence of observations (vectors) for each student and returns a sequence of estimated hidden states (GPA Level). In order to decrease the dimensionality of the problem, for the first, second, the third element in the grade number vector, we assign three if students have three or more courses with the grade of A, B, and C, respectively. Therefore, the first to third elements could get any values from 0 to 3, and the fourth and fifth elements are binary (0 and 1). 

%After initiating vectors for each student each semester, we convert each distinct vector into a single integer number and consider it an HMM observation. Going through students' transcripts, there are 216 different distinct combinations of vector grades; therefore, the HMM observation could be any number from 0 to 215. The model's output is the estimated academic-performance level that is an integer number between 1 to 4. Levels 1 and 4 corresponds to the lowest and highest academic performance, respectively. Given the estimated academic-performance level for each student each semester as a standard reference, we can track students' academic performances during their academic careers. As mentioned earlier, this academic-performance trajectory tracking helps university policymakers inform students at the risk of halting their studies and prioritize them for receiving support. \tabref{tab:HMM_input_output} presents the input and output of HMM for some arbitrary students. 

Based on an HMM model with 4 academic-performance levels and 216 possible grade-tuples, it is possible to provide examples of the desired input and output of the HMM model; see \tabref{tab:HMM_input_output}.  In the table, student number 1 has enrolled in four semesters and has four grade-tuples recorded in his academic history. In the first semester, he has no courses with A, B, or C grades as the first three digits are all zero. The fourth and fifth digits in the grade-tuple indicate that the student has grades DF and W in his first semester. His grade-tuple for the rest of the semesters is explained in the same manner.  The estimated academic-performance level for this student in all enrolled semesters is 1. As we see in the table, while students number 1 to 4 has a consistent academic-performance level during their academic careers, the academic-performance level for student number 5 has changed from 4 to 1 in his third semester. 

\begin{table}
\centering
\caption{Example students' grade-tuple sequence and corresponding estimated academic-performance level sequence}
\begin{tabular}{|c|c|c|} \hline
Student &Grade-tuple Seq.& Academic-per. Seq.\\ \hline
1&00011,10100,00211,00011& 1,1,1,1\\ \hline
2&11110,11200,21001,13000 & 2,2,2,2\\ \hline
3&22000,32000,30100 & 3,3,3\\ \hline
4&30000,31000,30000 & 4,4,4 \\ \hline
5&30000,30000,00010,00010 & 4,4,1,1\\
\hline\end{tabular}
\label{tab:HMM_input_output}
\end{table}

%% file: training.tex
\subsection{Initialization and Training}
For this particular problem, the process for initializing and training the HMM is non-trivial, especially given the total number of parameters that define the emission distributions (i.e., 4 states x 216 possible grade-tuple outputs).

First, we describe the process for providing an initial guess for the smaller 4x4 transition matrix, which has 12 linearly independent parameters (out of 16 total).  To identify an initial guess for HMM transition matrix, we analyzed students' transcripts to estimate how their GPA moves between the four levels identified in \tabref{tab:ClusterCGPA} -- this is equivalent to assuming a students GPA directly maps to an academic performance level (removing the randomness assertion stated earlier). The matrix below provides the resulting initial transition matrix (percentage scale):
\begin{equation*} \textrm{Initial guess for A}=\begin{bmatrix} 
81.81 & 17.10 & 1.02&0.07\\
7.90 & 78.43 & 13.52&0.15\\
0.74 & 8.54 & 84.44&6.28\\
0.34&1.00&10.32&88.34
\end{bmatrix}\end{equation*}
%In this matrix, rows 1 to 4 correspond to the academic-performance levels 1 to 4, respectively. As mentioned earlier in \tabref{tab:ClusterCGPA}, academic-performance level 1 has the lowest cumulative GPA while academic-performance level 4 is related to the highest cumulative GPA. For example, based on the numbers in the first row of the matrix, students in academic-performance level 1 stay at the same level with the probability of 81.81\%, and they move to academic-performance level 2, 3, and 4 17.10\%, 1.02\%, and 0.07\% chances respectively.   
As an initial guess for the vector $\pi$, we consider the distribution over students' academic-performance in their first semester. The matrix below shows the initial guess for $\pi_0$ (percentage scale):
\begin{equation*}\textrm{Initial guess for $\pi_0$}=\begin{bmatrix} 
20.14 & 19.07 & 33.50 & 27.29
\end{bmatrix}\end{equation*}
%This matrix shows that 20.14\% of FTIC undergraduate students at UCF start their education careers with academic-performance level 1, 19.07\% with level 2, 33.5\% with level 3, and finally 27.29\% with academic-performance level 4.
As we explained earlier, an HMM observation in this study for a student in a given semester is defined as a tuple with five elements. Elements 1 to 3 store the number of courses with grades A, B, and C, respectively, and elements 4 and 5 are binary variables, determining if the student has D/F and W grades in that semester. In order to find an initial guess for the emission matrix, we supposed that elements of the observed grade-tuples are independent of each other, each having a Poisson distribution (in the tuning process, the independence assumption between grade counts is removed). As such, for each academic-performance level, we can compute the probability of observing each grade-tuple, i.e., the grades counts in each tuple, using a joint but separable Poisson probability density function. Since there is a total of 216 different grade-tuple combinations in our data set, the initial emission matrix has four rows (corresponding to the four hidden states) and 216 columns (corresponding to the 216 observations). Due to space limitations, instead of showing the initial emission matrix with 864 parameters, we have computed the expected values (EV) of the counts for each grade across the 216 possible observations. The results are shown in the following matrix:
\begin{equation*} \textrm{Initial EV(grade)}=\begin{bmatrix} 
0.331&0.676&0.718&0.304&0.154\\
0.702&1.058&0.726&0.179&0.100\\
1.130&1.077&0.401&0.070&0.060\\
1.359&0.572&0.075&0.009&0.025
\end{bmatrix}\end{equation*}
In the above matrix, rows 1 through 4 correspond to academic-performance levels 1 to 4, while columns 1 to 5 provide the expected value for the number of courses with an A, B, C, D/F, and W grade each semester. For example, for students in academic-performance level 3 (the first row), in each semester, in the expectation, students earn 1.13 A's, 1.077 B's, and .401 C's. 

After initializing HMM parameters, the next step is to estimate their optimal value using the Baum-Weltch algorithm. The optimal estimates, obtained after 20 iterations for algorithm convergence, are listed below:
\begin{equation*} \textrm{A}=\begin{bmatrix} 
86.11 & 10.93 & 2.44&0.52\\
8.40 & 80.25 & 11.26&0.09\\
0.29 & 7.58 & 84.98&7.15\\
0.02&0.02&5.64&94.32
\end{bmatrix}\end{equation*}
\begin{equation*} \textrm{EV(grade)}=\begin{bmatrix} 
0.302&0.658&0.862&0.608&0.291\\
0.789&1.470&0.902&0.210&0.098\\
1.777&1.416&0.297&0.036&0.054\\
2.561&0.451&0.023&0.002&0.020
\end{bmatrix}\end{equation*}
\begin{equation*}\textrm{$\pi_0$}=\begin{bmatrix}
12.82 & 37.35 & 35.79 & 14.04
\end{bmatrix}\end{equation*}
\begin{equation*}\textrm{$\pi$}=\begin{bmatrix}
11.70 & 18.19 & 30.44 & 39.67
\end{bmatrix}\end{equation*}.
Again, instead of providing the probability of observing each of the 864 possible grade-tuples, the expectation of earned grades is reported.

%% file: analysis.tex
\section{Analysis}\label{sec:results}
The trained values of the HMM reported in \secref{sec:methodology} are of value in understanding academic-performance trajectories.  Based on the estimated transition matrix $A$, most students maintain their academic-performance level (mode) from one semester to the next. Looking at the diagonal of the matrix, 86.11\% of students in academic-performance level 1 in semester $t$ will remain at the same academic-performance level in semester $t+1$. These probabilities for students in academic-performance levels 2, 3, and 4 are 80.25\%, 84.98\%, and 94.32\%.  Based on these values, we are able to assert, with exception to performance-level 1, that increasing performance-levels are increasingly stable, so that a student in performance-level 4 is more likely to remain in the same performance-level as compared to a student in performance-level 2 or 3.

Matrix EV(grade) provides the expected values for grade counts using the estimated emission matrix. As illustrated in the matrix, students in academic-performance level 4 on average have more courses with A grades compared to the students in academic-performance level 1 (2.561 $>$ 0.302). Technically, while W grades are not used to compute semester GPAs, the last column of the matrix EV(grade) indicates there is a significant relationship between W grades and students' academic-performance level. Based on the columns, students with academic-performance level 1 are more likely to withdraw from courses in any given semester compared to students in other academic-performance levels (0.291 $>$ 0.098 $>$ 0.054 $>$ 0.02).

Matrix $\pi_0$ provides an estimate of the distribution of academic-performance levels for students' first semester. Based on the matrix, most students start their academic careers at academic-performance levels 2 and 3 (37.35\%+35.79\%). The matrix $\pi$, corresponding to the stationary probability distribution of the transition matrix A, reports the average probability students are in each academic-performance level, regardless of semester. The Comparison between $\pi_0$ and $\pi$ provides an indication of how the distribution of students' academic-performance evolves over time. For instance, it illustrates that students are more likely to be in academic-performance level 4 in later semesters compared to the first semester (39.67\%$>$14.04\%). Furthermore, the probability of having academic-performance level 1 in the first semester is more than the same probability for the later semesters (12.82\%$>$ 11.70\%). There are two potential reasons for the change in the distribution of academic-performance levels over time: first, since there is a significant correlation between students' first GPA and retention \cite{gershenfeld2016role}, students with low first semester GPA may halt, which causes the probability of students being in academic-performance level 1 to decreased for later semesters. Second, improvement in students' GPA over time could be another reason for the mentioned change in the academic-performance level distribution.
\begin{table}
\centering
\caption{Distribution over students' academic-performance level at UCF}
\label{tab:GPA_Level_Dist}
\begin{tabular}{|c|c|} \hline
Academic-per. level&Percentage\\ \hline
Level 1&11.75\% \\ \hline
Level 2&14.42\%\\ \hline
Level 3&19.10\%\\ \hline
Level 4&11.43\%\\ \hline
Other&43.30\%\\ \hline
\end{tabular}
\end{table}
\tabref{tab:GPA_Level_Dist} depicts the distribution over students' academic-performance level at UCF. Levels 1 to 4 correspond to the students who maintain a consistent academic-performance level over all enrolled semesters. The group \emph{Other}, corresponds to the students who have a transition between academic-performance levels during their academic career at UCF. The synthetic student number 5 in \tabref{tab:HMM_input_output} is one such example for this category of students. This student has an academic-performance level of 4 in his first two semesters and then switch to academic-performance level 1 in his third semester. As reported in the figure, 43.3\% of students change academic-performance levels during their academic careers.   

\begin{table}
  \centering
 \caption{Gender distribution for students with different academic-performance levels at UCF}
  \label{tab:Gender_GPA_Level}
 \begin{tabular}{|c|c|c|c|c|c|} \hline
    {}&Level 1&Level 2&Level 3&Level 4& Other\\ \hline
    Female& 8.7\%&12.8\%&21.0\%&14.3\%&43.2\% \\
    Male& 15.5\%&16.4\%&17.0\%&7.9\%&43.2\%\\ \hline
 \end{tabular}
\end{table}

\begin{table}
  \centering
  \caption{Race distribution for students with different academic-performance levels at UCF}
 \label{tab:Race_GPA_Level}
  \begin{tabular}{|c|c|c|c|c|c|} \hline
   {}&Level 1&Level 2&Level 3&Level 4& Other\\ \hline
    White& 10.7\%&14.1\%&19.8\%&12.7\%&42.7\% \\ %\hline
    Hispanic& 13.0\%&14.4\%&19.0\%&9.9\%&43.7\%\\ %\hline
    African-Am.& 16.1\%&17.7\%&15.0\%&4.9\%&46.3\%\\ %\hline
    Other race& 12.7\%&13.5\%&18.1\%&12.1\%&43.6\%\\ \hline
  \end{tabular}
\end{table}

\tabref{tab:Gender_GPA_Level} compares distributions over academic-performance levels for students with different genders. As the table suggests, female students are more likely to have a academic-performance level of 4 compared to male students (7.9\%$<$14.3\%). Also, female students have a academic-performance level 1 with less probability compared to male students (8.7\%$<$15.5\%)\footnote{Conducted chi-squared test demonstrates that the difference between academic-performance levels distributions for female and male students is statistically significant (\emph{p-value}=0).}. Based on these results, female students on average have a higher academic-performance level compared to male students, and gender seems to be an important factor affecting students' academic-performance levels at UCF. These results confirm previous findings obtained by Masic et al. \cite{mavsic2020relationship}, in which female students on average have a higher performance than male students. 

\tabref{tab:Race_GPA_Level} shows academic-performance level distributions for students with different races. As illustrated in the table, White students have the highest chances for acquiring academic-performance level 4, followed by Hispanic students and African-Am. students, respectively 
(4.9\%$<$9.9\%$<$12.7\%). Also, students of the White race are less likely to have academic-performance level 1 compared to Hispanic and African-Am. students (10.7\%$<$13.0\%$<$16.1\%)\footnote{Conducted chi-squared test demonstrates that the difference between academic-performance levels distributions for students with different races is statistically significant (\emph{p-value}=0).}. These findings can be supported by the U.S department of education \cite{de2019status}, which shows white students have the highest academic performance- in terms of degree awarded rate-, followed by Hispanic students, followed by American-Am. students. Therefore, similar to gender, race serves as a significant predictor for academic-performance level at UCF, which informs university policymakers to pay more attention to males and African-Am. students compared to other students since these students are considered at-risk students. 

Of particular interest to this paper is the \emph{Other} group, which consists of students whose academic-performance changes levels during their academic career. Among these students, 75.2\% switch their academic-performance levels only once, 20.4\% switch their academic-performance levels twice, and 4.5\% change their academic-performance levels more than 2 times during their academic career at UCF. The distribution over switching type for students with one switch is summarized in \tabref{tab:One_Switch_Dist}. Based on the table, the more common switches between academic-performance levels corresponds to level 2 \textrightarrow level 3 (27.8\%), level 3 \textrightarrow level 4 (27.7\%), and level 2 \textrightarrow level 1 (22.3\%). For students with one switch in their academic-performance level, 59.4\% improve their academic-performance levels, while 40.6\% worsen. For the group with two switches, the percentages of students ultimately improving and worsening their academic-performance level are 41.1\% and 15.7\%, respectively. Moreover, 43.2\% of these students go back to their initial academic-performance level after two switches (for example, level 3 \textrightarrow level 2 \textrightarrow level 3). 
\begin{table}
  \centering
  \caption{Distribution over switching type for students with one switch in academic-performance level}
  \label{tab:One_Switch_Dist}
  \begin{tabular}{|c|c|c|c|} \hline
    Switch&Ratio&Switch&Ratio\\ \hline
    level 1 \textrightarrow level 2& 2.4\%&level 3 \textrightarrow level 1&0.2\% \\ \hline
    level 1 \textrightarrow level 3& 1.0\%&level 3 \textrightarrow level 2&12.9\%\\ \hline
    level 1 \textrightarrow level 4& 0.5\%&level 3 \textrightarrow level 4&27.7\%\\ \hline
    level 2 \textrightarrow level 1& 22.3\%&level 4 \textrightarrow level 1&0.0\%\\ \hline
    level 2 \textrightarrow level 3& 27.8\%&level 4 \textrightarrow level 2&0.0\%\\ \hline
    level 2 \textrightarrow level 4& 0.0\%&level 4 \textrightarrow level 3&5.2\%\\ \hline
  \end{tabular}
\end{table}

\begin{table}
\centering

\caption{Comparing halt rate between students with no switch in academic-performance level and one switch academic-performance level}
%\textwidth
\begin{tabular}{|c|c|c|c|}
\hline
Staying in & Halt rate & Switching & Halt rate \\
\hline
\multirow{3}{*}{Level 1}&
\multirow{3}{*}{97.4\%}  
  & level 1 \textrightarrow level 2 &9.2\%\\
  &{}& level 1 \textrightarrow level 3 &6.5\%\\
  &{}& level 1 \textrightarrow level 4 &5.1\%\\
  \hline
 
\multirow{3}{*}{Level 2}&
\multirow{3}{*}{43.7\%}
  & level 2 \textrightarrow level 1 &73.6\%\\
  &{}& level 2 \textrightarrow level 3 &4.2\%\\
  &{}& level 2 \textrightarrow level 4 &$-$\\
  \hline

\multirow{3}{*}{Level 3}&
\multirow{3}{*}{23.8\%}
  & level 3 \textrightarrow level 1 &77.3\%\\
  &{}& level 3 \textrightarrow level 2 &8.7\%\\
  &{}& level 3 \textrightarrow level 4 &2.0\%\\
  \hline
 
\multirow{3}{*}{Level 4}&
\multirow{3}{*}{12.8\%} 
  & level 4 \textrightarrow level 1 &100\%\\
  &{}& level 4 \textrightarrow level 2 &$-$\\
  &{}& level 4 \textrightarrow level 3 &3.2\%\\
  \hline
\end{tabular}
\label{tab:Switching_halt_rate} 

\hfill
\end{table}

\tabref{tab:Switching_halt_rate} compares the halt rate for a subset of academic-performance level trajectories. Column \emph{Staying in} corresponds to students who keep a consistent academic-performance level. As the corresponding halt rate column suggests, the higher academic performance level, the lower the halt rate (97.4\% $>$ 43.7\% $>$ 23.8\% $>$ 12.8\%). Furthermore, we see that students that switch academic-performance levels from level 1 to other levels have a substantially improved halt rate (from 97.4\% to 9.2\%, 6.5\%, and 5.1\% for levels 2, 3, and 4). Also, switching from any level to level 1 increases the halt rate significantly. For example, all students whose academic-performance levels have changed from 4 to 1 left school without a degree. This percentage for students from level 3 and 2 to level 1 are 73.6\% and 77.3\%. However, there are surprising results for students starting at academic-performance level 3, where we see that students who worsen to academic-performance level 2 have a lower halt rate than if than students that had remained at level 3 (23.8\%$>$8.7\%). A similar pattern is observed for students whose academic-performance levels are changed from 4 to 3.  When combined, this evidence implies that switching from a high academic-performance level to a lower academic-performance level does not necessarily result in a higher halt rate.  Another surprising result emerges from the grouping and analysis of the academic trajectories that involve improving academic performance.  For students that improve their academic performance, their halt rate is substantially lower than those students that maintained a consistent academic performance level.  For example, the halt rate for students who transitioned from level 2 to level 3 is lower than those students who were always at level 3 (4.2\% $<$ 23.8\%). Even students that transitioned from level 1 to level 2 have a lower halt rate than those students consistently in level 4.   

%Finally, we investigate the impact of changing majors on switching academic-performance levels. From students who have changed their majors for once, 29.8\% have an increment in their academic-performance levels, 51.2\% have the same GPA level, and 19.0\% have a reduction in their academic-performance levels. Since the percentage for increasing academic-performance level is more than decreasing academic-performance level after switching major, changing major can be considered a motivating factor for students to improve their academic performance. In other words, students' interest regarding the major that they study has effects on their performance. 

%% file: conclusion.tex
\section{Conclusion and Future Research}\label{sec:conclusion}
In this paper, the authors analyzed students' academic-performance trajectories at University of Central Florida and examined their final academic outcomes in terms of halt rates. Unlike traditional statistical methodologies, our proposed approach is able to provide a standard point of reference for comparing student's GPA both across their enrolled semesters and across all students. Our proposed HMM model takes the sequence of course grades over multiple semesters and returns the sequence of estimated academic-performance levels. The estimated HMM parameters illustrate that while W grades are not involved in computing students' GPA, there is a significant relationship between students' academic-performance level and withdrawing from courses. Also, by tracking students' transitions between four main academic-performance levels, we observed that a significant portion of students (43.30\%), acquire a combination of academic-performance levels during their education.  Furthermore, in analyzing and comparing the halt rate for students with consistent academic performance levels, evidence suggests that students who constantly maintain a low academic-performance level are more likely to leave school without a degree. Meanwhile, for students with a change in their academic-performance level, it was shown that switching from any academic-performance levels to level 1 increases the halt rate substantially. Most surprising, however, was that for some academic-performance trajectories, when switching from a high academic performance level to a lower level, the halt rate did not increase but rather decreased.  In fact, the reduction in halt rate is even lower than if the student had remained at the same academic level. An investigation into the underlying reasons for such observations is left as a good avenue for future qualitative research. 

As discussed earlier in this paper, one of the main challenges of HMMs correspond to the dimensionality of the observations. In order to tackle this problem, two different approaches were used in this chapter. First, an upper limit was considered on the elements of the grade-tuple which restricted number of 
A's, B's, and C's to the maximum of 3 and transformed the number of D/F and W grades to binary variables. Second, since the probability of some grade combinations is zero, each unique grade-tuple was considered as one possible HMM observation. This assumption decreased the total number of unique observations to 216 (from 687). Although leveraging these problem-specific features helped reduce size of the problem that was addressed in this paper, in general such conditions may not always exist. The alternative approach to tackle this problem can be considered as a research question for future studies.

%% file: conference_101719.bbl
\begin{thebibliography}{10}

\bibitem{aghajari2020decomposition}
Z.~Aghajari, D.~S. Unal, M.~E. Unal, L.~G{\'o}mez, and E.~Walker.
\newblock Decomposition of response time to give better prediction of
  children's reading comprehension.
\newblock {\em International Educational Data Mining Society}, 2020.

\bibitem{bhattacharyya1946measure}
A.~Bhattacharyya.
\newblock On a measure of divergence between two multinomial populations.
\newblock {\em Sankhy{\=a}: the indian journal of statistics}, pages 401--406,
  1946.

\bibitem{boumi2019application}
S.~Boumi and A.~Vela.
\newblock Application of hidden markov models to quantify the impact of
  enrollment patterns on student performance.
\newblock {\em International Educational Data Mining Society}, 2019.

\bibitem{boumi2021quantifying}
S.~Boumi and A.~E. Vela.
\newblock Quantifying the impact of student enrollment patterns on academic
  success using a hidden markov model.
\newblock {\em Applied Sciences}, 11(14):6453, 2021.

\bibitem{boumi2021quantifying2}
S.~Boumi and A.~E. Vela.
\newblock Quantifying the impact of students' semester course load on their
  academic performance.
\newblock In {\em 2021 ASEE Virtual Annual Conference Content Access}, 2021.

\bibitem{boyer2011investigating}
K.~E. Boyer, R.~Phillips, A.~Ingram, E.~Y. Ha, M.~Wallis, M.~Vouk, and
  J.~Lester.
\newblock Investigating the relationship between dialogue structure and
  tutoring effectiveness: a hidden markov modeling approach.
\newblock {\em International Journal of Artificial Intelligence in Education},
  21(1-2):65--81, 2011.

\bibitem{de2019status}
C.~de~Brey, L.~Musu, J.~McFarland, S.~Wilkinson-Flicker, M.~Diliberti,
  A.~Zhang, C.~Branstetter, and X.~Wang.
\newblock Status and trends in the education of racial and ethnic groups 2018.
  nces 2019-038.
\newblock {\em National Center for Education Statistics}, 2019.

\bibitem{ebrahiminejad2019pathways}
H.~EbrahimiNejad, H.~A. Al~Yagoub, G.~D. Ricco, M.~W. Ohland, and L.~Zahedi.
\newblock Pathways and outcomes of rural students in engineering.
\newblock In {\em 2019 IEEE Frontiers in Education Conference (FIE)}, pages
  1--6. IEEE, 2019.

\bibitem{falakmasir2016data}
M.~H. Falakmasir, J.~P. Gonz{\'a}lez-Brenes, G.~J. Gordon, and K.~E. DiCerbo.
\newblock A data-driven approach for inferring student proficiency from game
  activity logs.
\newblock In {\em Proceedings of the Third (2016) ACM Conference on Learning@
  Scale}, pages 341--349. ACM, 2016.

\bibitem{garibaldi2012college}
P.~Garibaldi, F.~Giavazzi, A.~Ichino, and E.~Rettore.
\newblock College cost and time to complete a degree: Evidence from tuition
  discontinuities.
\newblock {\em Review of Economics and Statistics}, 94(3):699--711, 2012.

\bibitem{gershenfeld2016role}
S.~Gershenfeld, D.~Ward~Hood, and M.~Zhan.
\newblock The role of first-semester gpa in predicting graduation rates of
  underrepresented students.
\newblock {\em Journal of College Student Retention: Research, Theory \&
  Practice}, 17(4):469--488, 2016.

\bibitem{homsi2008hidden}
M.~Homsi, R.~Lutfi, R.~M. Carro, and B.~Ghias.
\newblock A hidden markov model approach to predict students' actions in an
  adaptive and intelligent web-based educational system.
\newblock In {\em 2008 3rd International Conference on Information and
  Communication Technologies: From Theory to Applications}, pages 1--6. IEEE,
  2008.

\bibitem{hovdhaugen2009learning}
E.~Hovdhaugen and P.~O. Aamodt.
\newblock Learning environment: Relevant or not to students' decision to leave
  university?
\newblock {\em Quality in higher education}, 15(2):177--189, 2009.

\bibitem{hull2005career}
E.~Hull-Blanks, S.~E.~R. Kurpius, C.~Befort, S.~Sollenberger, M.~F. Nicpon, and
  L.~Huser.
\newblock Career goals and retention-related factors among college freshmen.
\newblock {\em Journal of Career Development}, 32(1):16--30, 2005.

\bibitem{kaser2017modeling}
T.~K{\"a}ser, N.~R. Hallinen, and D.~L. Schwartz.
\newblock Modeling exploration strategies to predict student performance within
  a learning environment and beyond.
\newblock In {\em Proceedings of the seventh international learning analytics
  \& knowledge conference}, pages 31--40, 2017.

\bibitem{marbouti2016models}
F.~Marbouti, H.~A. Diefes-Dux, and K.~Madhavan.
\newblock Models for early prediction of at-risk students in a course using
  standards-based grading.
\newblock {\em Computers \& Education}, 103:1--15, 2016.

\bibitem{mavsic2020relationship}
A.~Ma{\v{s}}i{\'c}, E.~Polz, and S.~Be{\'c}irovi{\'c}.
\newblock The relationship between learning styles, gpa, school level and
  gender.
\newblock {\em European Researcher}, 11(1):51--60, 2020.

\bibitem{mirzaei2019modeling}
M.~Mirzaei and S.~Sahebi.
\newblock Modeling students’ behavior using sequential patterns to predict
  their performance.
\newblock In {\em International Conference on Artificial Intelligence in
  Education}, pages 350--353. Springer, 2019.

\bibitem{nguyen2013new}
L.~Nguyen.
\newblock A new approach for modeling and discovering learning styles by using
  hidden markov model.
\newblock {\em Global Journal of Human Social Science Linguistics \&
  Education}, 13(4), 2013.

\bibitem{pappas2016investigating}
I.~O. Pappas, M.~N. Giannakos, and L.~Jaccheri.
\newblock Investigating factors influencing students' intention to dropout
  computer science studies.
\newblock In {\em Proceedings of the 2016 ACM Conference on Innovation and
  Technology in Computer Science Education}, pages 198--203, 2016.

\bibitem{shapiro2018completing}
D.~Shapiro, A.~Dundar, F.~Huie, P.~K. Wakhungu, A.~Bhimdiwala, and S.~E.
  Wilson.
\newblock Completing college: A national view of student completion rates--fall
  2012 cohort (signature report no. 16).
\newblock {\em National Student Clearinghouse}, 2018.

\bibitem{shippee2011gpa}
N.~D. Shippee and T.~J. Owens.
\newblock Gpa, depression, and drinking: A longitudinal comparison of high
  school boys and girls.
\newblock {\em Sociological Perspectives}, 54(3):351--376, 2011.

\bibitem{tadayon2020predicting}
M.~Tadayon and G.~J. Pottie.
\newblock Predicting student performance in an educational game using a hidden
  markov model.
\newblock {\em IEEE Transactions on Education}, 2020.

\bibitem{asee_peer_34921}
L.~Zahedi, S.~J. Lunn, S.~Pouyanfar, M.~S. Ross, and M.~W. Ohland.
\newblock Leveraging machine-learning techniques to analyze computing
  persistence in undergraduate programs.
\newblock In {\em 2020 ASEE Virtual Annual Conference Content Access}, number
  10.18260/1-2--34921, Virtual On line, June 2020. ASEE Conferences.
\newblock https://peer.asee.org/34921.

\end{thebibliography}
